# Impactof interpersonal i nfluences on Employee engagement and Psychologicalcontract: Effectsof gua nxi, wasta, jeitinho, blat and pulling strings

Elizabeth Kassab Sfeir PhD
Université Saint-Joseph

elizabeth.kassab@usj.edu.lb

Abstract

This study puts forward a conceptual model linking interpersonal influences' impact on Employee Engagement, Psychological Contract, and Human Resource Practices. It builds on human and social capital, as well as the social exchange theory (SET), projecting how interpersonal influences can impact the psychological contract (PC) and employee engagement (EE) of employees. This research analyzes the interpersonal influences of Wasta in the Middle East, Guanxi in China, Jeitinho in Brazil, Blat in Russia, and Pulling Strings in England. Interpersonal influences draw upon nepotism, favoritism, and corruption in organizations in many countries. This paper draws on the qualitative methods of analyzing previous theories. It uses the Model Paper method of predicting relationships by examining the question of 'how do interpersonal influences impact employee engagement and psychological contract?' It is vital to track the effects of interpersonal influences on PC and EE, acknowledging that the employer can either empower or disengage our human capital.

# Introduction

The feat of survival at the organization, national and economic levels is dependent on human capital. In these competitive times, organizations need employees prepared to go above and beyond the call of duty. To do so, they must have empowered work relationships. Hence two psychological constructs, employee engagement and psychological contract, need to be simplified. With the dearth of management and social theories, researchers need to extrapolate further to demystify the impact of interpersonal influences on HR practices.

Interpersonal influences are a global phenomenon; hence businesses can improve their management practices by translating these concepts with a human resource lens. The areas of *Wasta (Middle East), Guanxi (China), Jeitinho (South America), Blat/Svyazi (Russia)* have been under scrutiny in the disciplines of anthropology, sociology, and ethnology. However, due to international business, the growth of the major economies, and the expansion of multinational companies in various regions, the importance of understanding and engaging in studies, including cultural features, have become vital (Velez-Calle et al., 2015). Managing an organizations' human resources in subsidiary operations is a requirement. Moreover, understanding an individual's native culture when the cultural distance between the manager's nation and the employee is pivotal (Hutchings & Weir, 2006). Even though there have been cultural studies on business practices, this research is intellectualizing the argument linking various interpersonal influences on Human Resource (HR) studies, a core foundation within any business/organization.

The purpose of this study is to provide a detailed framework illustrating the prominence of interpersonal influences over business settings. As Velez-Calle et al. (2015) infer, even though business practices are motivated by economic interests, it is naïve and inappropriate for the business practitioner to dismiss the influence that culture has on determining certain business practices making processes and outcomes of business relationships and negotiations. Hence, this work aims to deconstruct the forces of interpersonal relations and provide a critical lens on the impact of employee engagement and psychological breach of contract. It examines the research

question "How do interpersonal influences impact employee engagement and psychological contract?"

Background

According to human capital theory (Youndt & Snell, 2004) cited in Aklamanu et al. (2016), companies resource their human capital from individuals in the labor market through HRM practices. When businesses hire people with specific knowledge, skills, and abilities (KSAs), they invest in their workers and create their human resources through training and career growth. As a result, HR practices cultivate the Human Capital (HC) needed (Aklamanu etal.,2016).

The author Becker (1964), cited in Tomer (2016), refers to Human capital (HC) as the "resources in people." The greater the skills and knowledge that people have, the higher monetary and psychic income received. Tomer (2016) notes that Becker's definition is in line with mainstream economics, where Human Capital fits within the neoclassical analysis framework. This definition is an essential concept in standard HC theory, where HC investment contributes to noncognitive human development and changes to human relationships. Becker's research reiterates the importance that human relationshipsha ve on increasing HC formation.

Social capital, which refers to social relationships and their contribution to economic and social institutions, is included in the definition of human capital. According to various scholars, such as Tomer ( 2016) and Coleman (1990), social capital (SC) is a social resource that enables actors to achieve their end goals and encompasses families, organizations, communities, and societies as a whole, as well as, within each person.

According to Social Exchange Theory (SET), person-to-person "interdependent and contingent" exchanges are the foundation for all societal transactions and relationships (Cropanzano and Mitchell, 2005, p. 874; Emerson, 1976). One of the SET's core assumptions is that different social interactions are focused on reciprocity, psychological contracts, and other forms of social interaction. As a result, SET is an appropriate foundation for how intra-organizational social networking infrastructure affects capabilities and knowledge management by shaping employees' attitudes toward collaboration and organizationalcitizenship ( Oparaocha, 2016).

Individuals are mainly involved in social exchanges and social capital acquisition for personal gain (whether that gain is formal, such as job-related favors, or informal, such as friendship and favors outs) (Oparaocha, 2016). These friendships and favors are found in many cultures' interpersonal influences to advance people's careers and give them prominence and accessibility in the workplace.

This research adds to the body of knowledge in several ways. First, we advance interpersonal factors to demonstrate their significant effect on HR practices, a novel contribution. Second, we can create the connection explaining the psychological impact of these "interpersonal relationships" by analyzing the social exchange theory and the principle of psychological contract fulfillment within a social exchange mechanism. Lastly, we show that psychological contract fulfillment plays a mediating role in the social exchange relationship, contributing to the operationalization of the social exchange relationship in any organization (Tekleab, Takeuchi, & Taylor, 2005) cited in Birtch et al., (2016).

Unlike a written work agreement's explicit terms and conditions, a psychological contract reflects employee beliefs and assumptions regarding implied commitments and responsibilities in an employment exchange between the employee and the employer (Rousseau, 2001). Simply put, it is a worldwide cognitive assessment of how well past commitments have been kept (Rousseau, 1990, 1995). Since a job exchange consists of both concrete (transactional) and intangible (relational) components, employees shape cognitions regarding the inducements and incentives (such as pay, wages, preparation, and development) they obtain from employers for their work effort and contributions (Birtch et al., 2016).

LiteratureR eview

HR practices impact the employee's attitudes and behavior at work (Aggarwal et al., 2007). They delineate two psychological concepts that achieve outcomes, employee engagement (EE) and psychological contract (PC). The similarities between the two are that they are both mechanisms for motivation. They are both considered constructs of employment relationships. Various interpersonal influences (s) like *Wasta* in the Middle East and *Guanxi* in China indicate a connection between the individual and the organization. Thus, it is relevant to examine how EE and PC might play a part in HR practices in various cultures. Chang, Hsu, Liou, and Tsai (2013) explain two types of psychological contracts: transactional and relational. They are on the opposite end of the continuum; hence, these contracts affect innovative behavior. Aggarwal et al. (2007) state that the construct PC refers to the "employee expectation from and towards his/her job/organization"; "employee engagement also operates in the context of employee and his/her role and the organization"(p.317).

Aggarwal et al. 2007 cited in Rousseau (2001), define PC as a "set of expectations held by the individual employee that specifies what the individual and the organization expect to give and receive in the working relationship" (p.314). The authors infer that EE and PC are social exchanges. They are duly explained by the social exchange theory (SET) that suggests that, although formal contracts and economic issues drive the relationship in an organization, a social element evolves. Aggarwal (2014) states that psychological contracts involve many unspecified obligations that cannot be discussed in courts of law, as a psychological contract is a social exchange relationship. A core aspect of the employee-employer relationship is the psychological contract (PC), which has become in itself an established area of research (Persson & Wasieleski, 2015). PC tries to capture the relationship between employee and employer, which is constantly changing.

Most human resource departments consider employee engagement and commitment high on the agenda (Bal, Kooji & De Jong, 2013). Most research has shown that having a committed workforce can decrease turnover and increase organizational performance. Bal et al., 2013, state that it is essential to attract and sustain staff members' engagement levels due to the aging workforce and the difficulties inre taining talent.

Aggarwal et al. (2007) explain that an organization must examine the challenges that face an organization in managing talent, exclusively that the global environment is evolving and challenges increase (figure 1 below). Organizations that are effective, innovative, and make a significant difference in revenue examine their employees' current talent pool. Ryde (2010) states thats taff motivation is no longer popular, but employee engagement is.

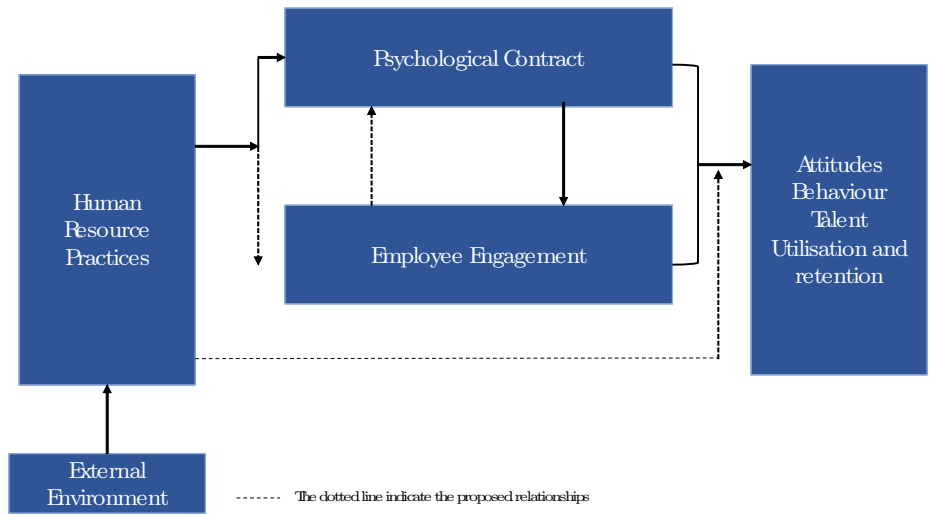

Figure 1: Illustrative framework by Aggarwal et al., 2007, p.322

Social exchange theory (SET) is a known framework that discusses the psychological contract process. In a simple formula, where an individual does someone a favor, there is an expectation of something in return in the future. This is also applied to interpersonal influences. In various definitions below, an interpersonal influence is about what can be gained and received from the other counterpart. Seeing that the relationship is very similar, interpersonal influences are a part of the SET theory. It is an explanatory theory and has direct applications in predicting employee and employer relationships; when employees receive economic and socio-economic resources from a company, they feel they need to repay it. The concepts of trust and reciprocity are, thus, the main features of SET. This theoretical approach is also the foundation for employee engagement (Aggarwal et al., 2007, Aggarwal, 2014). One of the ways that employees can repay is by having an increased level of engagement.

Psychological contract breach (PCB) has significant results at the individual, group, and organizational levels, according to Ahmed and Muchiri (2013). They also state that PCB also has links to job satisfaction, job neglect, organizational citizenship behaviors, the intention to leave, and the employee's well-being. Ahmed and Muchiri (2013) explain that researchers have become increasingly interested in PC and its impact on employees over the last twenty years. PC is negatively related to employees' trust in management, job satisfaction, talent retention, employee performance, citizenship behavior, and employee commitment.

The author Kahn (1990), cited in Kwon & Kim, 2019, defines personal engagement as an individual's psychologicals tate where employees give their 'full personal selves 'and invest various energies into their work. Theseenergies are physica l, cognitive, and emotional, and the amount ofe nergy they invest also leads to different outcomes (Kwon & Kim, 2019). Schaufe li, Salanova, González-Romá, and Bakker (2002) expounded on Khan's concept and further defined it as "apositive, fulfil ling, work-related state of mind that is characterized by vigor,dedication, and absorption"(K won & Kim, 2019, p74).

The information provided by Bhatnagar and Biswas (2010) is essential knowledge regarding relationships and social exchanges in the workplace, and thus, it is highly relevant in the current examination of interpersonal influences and their impact on individuals. There is currently little literature regarding employee engagement and interpersonal influences. This is an important area to explore to understand the influences that affect employees' psychological engagement in their work.

Arrowsmith and Parker (2013) state that 'employee engagement (EE) is a crucial element of human resource management (HRM). They explain that the EE involves the employee's enthusiasm for the organization and what the job involves. Arrowsmith and Parker (2013) explain that an engaged employee has a high degree of affective and cognitive commitment. This means that 'they go the extra mile' in exercising discretionary effort (Daniels, cited in Arrowsmith & Parker, 2013). Rayton and Yalbik (2014) highlight the fact obligations are likely to be more engaged and will not leavethe organiza tion quickly.

Kahn (cited in Aggarwal et al. 2007) defines EE as the "physical, cognitive and affective involvement of an employee and the harnessing of organization members' selves to their work roles" (p.316). Employee Engagement (EE) is the degree to which individuals are active and absorbed in their roles. It is the emotional involvement of an employee in his work role. Bhatnagar and Biswas (2010) state that Employee Engagement is the key to retaining talent. They also state that those employees who experience well-being within an organization have a lower intention to quit. This factor provides a competitive advantage to the organization. The information provided by Bhatnagar and Biswas (2010) is essential knowledge regarding relationships and social exchanges in the workplace, and thus, it is highly relevant in the current examination of *interpersonal influences* and their effects on individuals.

HR processes and practices play a significant and pivotal role in relationships between employee and employer. They also play an essential role in shaping the employees' PC and EE. Aggarwal et al. (2013) and Aggarwal (2014) confer that there have been a significant number of studies that link HR practices (HRP) and PC. This statement is pivotal to this research. Interpersonal influences, as stated before, are a relationship between the patron and the individual, where there is an attempt to gain privileges or resources.

Interpersonal influences are a part of everyday global business. Hence understanding the day-to-day transactions (*Wasta, Guanxi,* etc.) and how they impact employees is essential. Below are a few examples that support the impact of interpersonal influences on employees.

As inferred by Sfeir (2019), the data confirms that *Wasta* is a part of the organizational culture within universities. Over 20 interviews with employees show how they feel about *Wasta* and its influence on their working relationships. The author also reports that employee's motivation is affected and their job satisfaction. Furthermore, creating the *Wasta* organizational cultural model (WOC) illustrates how employee engagement decreases due to this interpersonal influence. This model shows that the individual relationships between people that work with each other are based on either negative or positive associations.

These informal networks in emerging markets are perceived as suspicious. This is due to their moral ambivalence and association with cronyism, cliques, and other negative phenomena (Horak et al., 2020). Moreover, societies do not take it as seriously as they believe that informal networks will become superfluous after organizations become more formalized. However, research must concentrate on reforming formal institutions and capacity building for transparency, good governance, and sustainability (Horak et al., 2020).

The most common perception of informal networks suggests that social networks can help deal with formal limitations. This is frequently referred to as mutual support or personal trust in the language of participants. The situation is significantly more complicated in the words of commentators. According to the 'topographical map' of existing perspectives on social networks, scholars make choices implicitly or overtly. First, they categorize 'nodes and ties' as either personal (represented by humans) or impersonal (defined by organizations). Second, they examine networks internally (exploring their composition and features) or externally (exploring their implications). (Ledeneva 2006) cited in Horak et al., 2020)

In terms of social capital, these ties are linked to the benefits and opportunities that people gain because of their membership in specific communities or networks (Horak et al., 2020). Other research reveals that the guanxi effect is prevalent and persistent among Chinese managers (Bian, 2018, 2019) cited in Horak et al., (2020) and among expatriate managers working in an increasingly globalized China. However, this study provides a timely analysis of the importance of culturally specific social capital in expatriate career development. It also focuses on understanding the informal networks and cultural intelligence that facilitates global organizations in various locations (Horak et al., 2020).

Except for Guanxi, international business studies have generally ignored many other informal networks as a research subject. However, it would be difficult for theory development to advance further should interpersonal influences not be considered. Other than studies on Guanxi, the current level of knowledge on informal ties in international business is limited to svyazi (Russia), yongo (South Korea), jinmyaku and gakubatsu (Japan), Wasta (Arab countries), jaan-pehchaan (India), and jeitinho (Brazil)(Horak et al., 2020).

The 'exchange' process between HRP and PC begins at the recruitment and selection process level and continues through the organization's life. Interpersonal influences *Wasta, Guanxi, Jeitinho, pulling strings, and Blat* also begin. Researchers infer that a firm's innovation is embedded within its social context (Jiang & Liu, 2015). They further indicate that the employee's behavior is based on the inter-social relationships within the organization. This relationship relies on the social capital perspective, that effectiveness is not based on individual performance but a vital social context.

The relationship between employees and their organizations is one of exchange (Richard et al., 2008). This exchange is an essential theoretical point showing the relationships' impact on organizational effectiveness and performance. This exchange is linked to interpersonal influences, like *Wasta*, *Guanxi*, *Jeitinho*, and *Blat*, to the theory of transaction costs. This theory is defined as "any activity which is engaged in, to satisfy each party to an exchange that the value given and received is in accord with his or her expectations" (Ouchi, 1980: 13) cited in Richard et al. 2008, p. 818. Moreover, the authors describe that transaction cost analysis (TCA) examines three modes of exchange: markets, bureaucracies, and clans. Furthermore, HRM researchers have acknowledged the 'role of context,' which is the setting that surrounds the 'obligations of the psychological contract' that aids in the workplace relationship changes (Kutaula, 2019).

Smith et al. (2012) discuss the interpersonal influences *of Wasta, Guanxi, jeitnho,* and *svyazi* within organizations as essential and culturally distinctive. These four principal modes of interpersonal influences are significant in understanding their effects on HR processes; more specifically, the comparison of interpersonal influences from a range of other cultures with the phenomenon of *Wasta* is also instructive. Some scholars are interested in comparing exclusively within individualism-collectivism, dimensions of cultural variations, Torres et al. (2015). It is crucial to examine the interpersonal influences due to the shift in global power—moreover, its impact on employeeengagement and the psychologic al contract.

There is little literature regarding *Jeitinho, svyazi*, and *Wasta*. The available literature examined below shows the similarities and differences between similar yet different cultural phenomena. There is little more information regarding *Guanxi*, which is not surprising due to China's global competition in our business environment. The interaction of social networks and corruption is dependent on the institutional environment, which is very different in China and Russia. Corruption is seen as a reciprocal exchange within a social network based on its culture and society. These national structures include the nature of morality and trust, which is the pillar for relationships in different networks(W illiams& Bezeredi, 2017).

These practices are essential to examine becausethey prevent t he emergenceof a meritoc ratic process and encourage the persistence of nepotism and corruption in countries.H ence, social capital's "dark side"uses t hesepersonal connect ions to bypass formal processes and procedures. (Williams & Bezeredi, 2017).

**Interpersonal Influences**

*Guanxi*, which is Chinese for 'connections,' is deemed necessary for interpersonal relationships within the Chinese culture (Smith et al., 2012). It is known as an informal personal connection between two people. Hutchings and Weir (2006) describe *Guanxi*as Chinese personal connections based on solid family networks and Confucian ethics. Hutchings and Weir (2006) describe *Guanxi* as a relationship between two people that encourages individuals "to give as good as they get" (p.143). A Chinese national will turn to their Guanxiwang, or network, for help. Trust, which is central to *Guanxi*, is called xinyong, and this is core to business activity, with the 'consultation' being the main focus. *Guanxi*refers to an individual 'drawing on connections' to secure favors within personal relationships. It is an exclusive network where the Chinese propagate with energy and imagination (Torres et al., 2015). The authors attest that *Guanxi*is about mutual obligation, understanding, and assurances. It also governs the attitudes of Chinese people in long-term social and business relationships.

So basically, it is the interpersonal linkages with ramifications of continued exchange for favors. Torres et al. (2015) ascertain that *Guanxi*has a practical application in management. In establishing *Guanxi,* several characteristics arise, including direction of power relations (whether vertical or horizontal), the level of intimacy, the uniformity of goals, and the willingness to maintain harmony (Hwang, 1997 cited in Torres et al., 2015. Nitsch and Diebel (2007) discuss that individuals who cannot achieve their goal directly need a third party involved. This third person who shares *Guanxi* with both sides can be a mediator for the person requesting. The authors describe five steps used; first, they gain face; second, draw connections, third estimate their benefit, fourth they see how much emotion is involved and finally, decision making.

The Brazilians use what is known as *jeitnho*, defined as "little way out or adroitness" (Smith et al., 2012, p.336). This term is used frequently by Brazilians in both social settings as well as business organizations. Initially, it was used with football players and referred to creative ingenuity in achieving quick results to problems. Football players used this interpersonal influence to work around bureaucratic rules or avoid potential issues with superiors in high positions. Duarte (2011) expounds the concept of *Jeitinho* through charm, defined as the 'power to attract, please and fascinate which creates positive rapport in social interaction' p. 29. Charm is a highly respected characteristic in Brazilian society. This characteristic is socially intertwined with *simpatia*, which can empathize and work towardsha rmony in interpersonal relations.

*Jeitinho* is a solid behavioral characteristic in many business organizations. Torres et al. (2015) state that is short-term solutions to problems. It is a vital characteristic that does not require previous relationships between the parties. They are usually anonymous figures that are only needed depending on one's necessity and the other individual's power to grant it. Torres et al. (2015) explain that achieving *Jeitinho* includes ' ways of circumventing bureaucratic rules or methods of handling potential difficulties with superiors in a robust hierarchicalcontext' p. 80.

Within the Brazilian society, Barbosa (cited in Torres et al., 2015) describes the *Jeitinho* on a continuum with both extremes of positive and negative, where the extreme positive approaches constitute a favor-like action and the negative extreme gives the idea of corruption. There are two types of *Jeitinho*: dar um *Jeitinho,* which means having an out, and *Jeitinho* Brasileiro, meaning the little Brazilian way or the Brazilian way of doing things (Nes, 2016). The first type suggests there can be a solution to the problem no matter what the situation is, and the second type has a negative connotation where it is close to corruption. This second term uses creativity more and is a part of everyday Brazilian life. This type of phenomenon is explicit and authentic and used by everyone. Torres et al. (2015) explain that management uses *Jeitinho* at different organizational levels, which is more of magnitude than incidence. *Jeitinho* either opens up to corruption or establishes social justice.

*Svyazi* or *Blat* is a Russian term similar to *Guanxi* attest Batjargal and Liu (cited in Smith et al., 2012). A more widely used, the informal procedure is called *Blat*, known to be the overt use of corruption, states Ledevena (cited in Smith et al., 2012). The authors declare that *Blat* is a specific form of *svyazi*. They infer that it became more frequent in the period of the Soviet command economy. The *svyazi* network is less personalistic than *Guanxi*. Job recruitment's 'dark side is known as '*blat*' (Onoshchenko & Williams, 2013). The authors discuss *Blat* during the soviet command economy, where having friends was strategic and imperative, as the issues were not about money but the shortage of goods and services. Access to these goods and services was through personal connections. So, the wide variety of connections was significant in times of need. A common phrase during that time was 'it is better to have a hundred friends than a hundred rubles' (Onoshchenko & Williams, 2013, p. 256).

*Blat* is used to negotiate many aspects of life, including requiring foods, holidays, schools, and universities to medical services. This type of influence is looked upon favorably as it helps to cope with the difficulties in life. These concepts are in the Soviet world, where there is no need for direct compensation (Onoshchenko & Williams, 2013). Ideas and perspectives have changed regarding personal networks. Making a living and getting a good job is now critical in the marketplace. It now seems that these connections are used to access employment, loans, and educational places, as Ledeneva (2013) cited in (Onoshchenko & Williams, 2013). Hence, the concept is access to assets. It is not about giving friendly help but being reimbursed in the form of money. Al Rahami (cited in Onoshchenko & Williams, 2013) also infer that this is also the case with *Wasta*.

Consequently, *Blat* is now considered more 'materialized,' losing the human warmth and having a negative meaning. This concept is especially the case with the younger generations. Onoshchenko & Williams (2013) infer that *Blat* is a form of corruption, where public office is used for personal advantage. The authors also attest that personal connections are vital to getting suitable employment. However, the authors note that there is minimal research on job recruitment.

*Pulling strings* is a British concept that is an idiomatic phrase. Smith et al. (2012, p. 337) state that it refers to "obtaining favors, particularly through links with influential persons." These 'links' are from time-honored relationships that come from family connections or shared schooling (in the American context, Ivy League, or, in British terms, the Old Boys' Club), or are developed from short term, 'by chance,' contacts. There are very few scholarly articles written on the subject through researching the topic online, except for *Guanxi* and *Wasta*'s comparison. Regarding getting employment, *Guanxi* is disapproved of in Western circles, where formal procedures are more equitable and fairer. Onoshchenko & Williams (2013) also reiterate the lack of information regarding *'pulling strings,'* widely used in the English-speaking world. They state that no empirical research has been conducted on the influence of this practice. However, the authors cite Smith et al. (2012) to reveal that the English people's attitude towards pulling strings is more favorable than the Arab, Chinese, Brazilian, or Russian world.

The term *Wasta* is known in the Middle Eastern countries as a type of social capital, however, spoken negatively (Bailey, 2012). As the author states, in the United Arab Emirates, *Wasta* implies that a person obtains or favors another person. Authors like Bailey (cites Cummingham & Sarayah (1993) and Zahra (2011) in discussing *Wasta* as being the predominant factor in receiving a job, which then creates disdain and much controversy. Religions like Islam play a role in *Wasta*, where it is seen to be negative, and it is condemned and is known to be socially unfair.

'*Wasta* involves a social network of interpersonal connections rooted in family and kinship ties and the exercise of power, influence, and information-sharing through social and political-business networks' (Iles et al., 2012, p. 472). Mohamed and Mohamad (2011) further explore how *Wasta* plays a significant role in hiring, selecting, and promoting employees in the Arab world. Of course, *Wasta* is compared to the Chinese *Guanxi* – as previously stated, a social network based on Confucian ethics. The author states that *Guanxi* may be beneficial for an organization's competitiveness and performance. However, this does not necessarily apply to the situation vis-à-vis *Wasta*. Many have blamed *Wasta* for the Arab World's poor economic performance and draining of financial and human capital. They also state that *Wasta* has become an issue for those seeking it, those granting it, and the government. Ahmed and Hadia (2008) cite Makhoul and Harrison (2004), who characterize *Wasta* as inefficient and have warned that it could lead to poor job performance and economic decline. They also suggest that having *Wasta* will feed injustice and frustration to those qualified for the job but do not have *Wasta*. Kassab (2016) illustrated the adverse effects that *Wasta* has on HR practices. The researcher's results showed that over 58.6% of people obtained their job through a *Wasta*.

**The objective of the study**

This study aims to deconstruct and uncover the consequences of interpersonal influences on the organization's employees. It exposes the dilemmas that are encountered when interpersonal influences are used in negative ways. How do interpersonal influences impact employee engagement and psychological contract?

These authors state that HRP notably impacts employees' development and the PC they fulfill (Bal et al., 2014). Through all the steps of being recruited, having a performance appraisal, reviewing their benefits package, or receiving recognition for what they have done, the employee will interpret all these, influencing how they react within the organization (Chang et al. 2013). Should the company fail to do so, that is, fulfill their obligations (what the employee perceives), this will detract from the employment relationship and employees changing what they feel obligated to offer. These authors have been able to conclude that HRP is an essential factor in influencing the PC. There still needs to be more work done on the PC and EE and their relationship. Aggarwal et al. (2007) state that, even though there is research on the impact of HRP on organizational outcomes, there is minimal information on the effects of HRP on individual employees.

PCB occurs when the individual recognizes that the company did not fulfill one or more of the promises made. Much of the research on PCB discusses how employees react to employer breaches, and what is found is that employees respond negatively. This reaction decreases job satisfaction, which reduces an employee's contribution to the exchange context and effectively rebalances the relationship. Rayton and Yalabik (2014) infer that this work engagement is less likely to occur when employees feel that their firm meets their obligations and when the employees are happy with their jobs. It is an extension of PC that isoffic ially entitled PBC.

Methodology

The researcher builds on qualitative research as a collection of interpretive and material acts that make the world visible. Qualitative researchers investigate phenomena in their natural contexts, aiming to make sense of occurrences in terms of people's meanings (Denzin & Lincoln, 2011). This research project views the details through a social constructivism/interpretivism paradigm. Hence, "they are not simply imprinted on individuals but are formed through interaction with others (hence social construction) and through historical and cultural norms that operate in individuals' lives" (Creswell, 2015, p.67). The writer is basing this research on inquiry, presenting a series of logically related steps. This research is not based on quantitative statistics but the gathering of ideas and building a stronger foundation.

| Objective | Research Question | Hypotheses | Independent Variables |
|---|---|---|---|
| To develop Aggarwal's Integrative framework on Employee engagement and psychological constructs to include interpersonal influences. | How do interpersonal influences impact employee engagement and psychological contract? | $H_0$- Interpersonal influences affect employee engagement $H_1$- Interpersonal influences do not affect employee engagement $H_0$- Interpersonal influences affect the psychological contract $H_2$- Interpersonal effects do not affect the psychological contract | Employee engagement Psychological Contract |

Table 1 - Research Objectives and Hypotheses

The method used for this research is known as the model paper. The model paper aims to create a theoretical structure that predicts relationships between concepts (Jaakkola, 2020). A conceptual model can define an occurrence, an object, or a process and demonstrate how it operates by revealing antecedents, consequences, and contingencies specific to the construct (s) being examined. This entails a type of theorizing in which the aim is to build a nomological network around the central concept, using a systematic approach to analyze and detail the underlying links and processes at work (Jaakkola, 2020).

This study examines the current PC and EE theories in an illustrated framework, revealing the impact of interpersonal influences. By building the theoretical framework with human and social theory and understanding the nature of social exchanges, various nodes are added to conceptualize the effects on employees.

As Jaakkola, 2020, infers, a model paper identifies previously unknown relations between constructs, adds new constructs, or illustrates why some process elements result in a specific outcome. This theoretical method adds to existing information by outlining an entity: its purpose is to detail the entity and its relationship to others as per author MacInnis, 2011, cited in (Jaakkola, 2020). In a philosophical paper, the researcher's imaginative reach is unrestricted by evidence-related constraints, enabling them to investigate and model new trends where little observational data are accessible (Yadav 2010) cited in Jaakola, 2020.

Implications

The interpersonal processes of influence discussed above – *Wasta, Guanxi, Jeitinho, Svyazi, and 'pulling strings'* (Smith et al., 2012) – have a joint base being interpersonal linkages that are not formal. Their differences are in the relationship's hierarchical nature and how intense and long the relationship may last. As the abovementioned authors note, *Guanxi* is based in Chinese business and social activities, whereas, *Wasta* is used in every critical business decision. In terms of similarities, *Guanxi* and *Wasta* occur in a more hierarchical relationship that comprises a long-term and emotional commitment. Obtaining *Wasta* means that there has to be loyalty given to that person who has helped get the 'position or favor' for another (Mohamad & Mohamad, 2011). So, in other words, power is given to that person who has successfully helped the other. It means that the person is 'beholden" to the person who is their *Wasta*. In Chinese culture, the failure to sustain a *Guanxi* relationship implies a loss of face; in a similar vein, when a person receives a favor through *Wasta*, it involves the continuing obligation to uphold the person's honor granting the privilege. *Wasta* has a social influence on the person's relationship at work – hence the notion of 'the weak, get weaker, and the strong, get stronger.

*Wasta* is a pervasive and pernicious feature of the Middle East culture, and its effect on performance and standards within organizations in the region is unfortunate. Though Wasta may never be eliminated from business practices, it challenges future researchers as it is so endemic to the culture. Researchers need to find ways to counter and negotiate it while introducing new and healthier HR practices to ensure optimum performance in workplaces across the region. In doing so, it will help the future leaders of this troubled region emerge.

Figure 2: EE & PC with Interpersonal Influences - *updated & adapted by the researcher*: based on the framework by Aggarwal et al., 2007.

As PC refers to the "employee expectation from and towards his/her job/organization,"; employee engagement as Social exchange theory (SET) is a known framework that discusses the psychological contract process. In a simple formula, where an individual does someone a favor, there is an expectation of something in return in the future.

In this framework, Interpersonal Influences have been placed as a node within, indicating that it also affects HR practices, affecting the employee's psychological contract and engagement. It is imperative to illustrate the impact that a country's national culture has on business practices. In addition, the researcher added the macro-environmental factors as a node.

Hence, the figure above shows that should there be an interpersonal Influence involved within the organization's HRPs, (indicated by the arrow connecting from the box Interpersonal Influences to HRPs), then HRPs also affect the employee's PC. It is crucial to understand that what is told and promised to "an employee" is taken seriously and, in turn, affects the employee's engagement with the organization. The figure also illustrates the micro and macro-environment surrounding the HRPs, including the social, legal, technological, and cultural phenomenon. The employee's motivation and satisfaction are also measured, which affects their attitudes and behavior, and retention within the organization. The addition of Interpersonal Influences is vital to recognize within the PC and EE model. It shows its dual effect on both the organizational culture and individual employees. Hence, this model's addition is an important avenue to pursue within PC and EE and its impact on individuals. The broken arrows from PC and EE to the interpersonal influences illustrate that PC and EE are independent variables and interpersonal influences are dependent variables. The links indicate that interpersonal influences have an impact on HR practices and EE and PC. Thus, signifying that $H_0$ is valid and interpersonal influences have an impact on EE and PC.

Moreover, therei s ane ed to link interpersonal influences in social capital theory and SET. Bailey (2012) notes that social capital is imperative to community development and improving economic health. This builds community alliances, allowing people to establish a sense of connection, build membership, and share work on civic projects. It also helps build trust. Once trust is built between members, favors are exchanged, confirming that each person will do what is needed for the other. Once these favors are defined within "interpersonal influences," it allows for a realistic image of organizational culture. As Bailey states, "true social capital is a dialectical exercise built on trust and reciprocity" p.2. Hence, building further on social capital theory, interpersonal influences on HR practices have a pivotal effect on the company's community alliances, membership, and organizational culture.

Velez-Calle et al. (2015) state that interpersonal influences can be seen from four different perspectives. These perspectives are corruption, a different path to weak institutionalism, social capital, or derived from specific cultural dimensions (p. 284). The author asserts that the first dimension distinguishes itself by emotional decision-making, which is dynamic and not logical. This would lead to economic weaknesses and inappropriate allocation of resources/and or choice of adequate partners (Velez-Calle et al., 2015). The second perspective focuses on a descriptive explanation of networking used to surmount the weak institutionalization within a country. When there is instability in the country's legal framework and very little formal support from formal institutions, businesses would eventually establish certain connections to facilitate operations (Velez-Calle et al., 2015). According to the authors, the third perspective looks at these practices as social capital. The development of business opportunities, based on personal connections, is not within one society but is a widespread global practice. As the authors explain, the last perspective indicates that networking is the process of specific cultural dimensions; *Wasta* often sways the situation for those trying to gain positions in the workplace, receive better compensation, and be promoted in public sector organizations. In this way, it is also similar to 'piston' in North Africa, as Iles et al. (2012) pointed out.

**Conclusion**

Further exploring into the evolution of interpersonal influences on Human Resource Practices is imperative. It posits rationales for its use in these societies. A better understanding of the relationships invokes improved human capital development. Human beings are amazing creatures, and, despite the best efforts, no research has succeeded in fully explaining human behavior's intricacies. As our world has become ever more diverse and multicultural, understanding human behavior has become one of the most critical aspects of personal and professional life. Kipkebut

(2010) states, "various studies have stressed the benefits to organizations of a loyal and committed workforce." Therefore, any organization that wishes to maintain a healthy organizational culture must understand the human capital that works within it.

Cultural practices in HR and HRM have clear ramifications for adopting globally recognized and ratified management practices in any sector. Thus, the identification of such global practices is central to the present study. Taking Aggarwal's model of HR practices and its connections to the EE and PC constructs, and linking interpersonal influences, provides organizations with further knowledge of the impact these "international business "exchanges have on employees and leadership. Through the research conducted, the researcher has been able to link the effects of *Wasta, Guanxi, Jeitinho, and Blat* on the psychological contract and employee engagement. When a person believes that they should receive certain benefits and does not, this creates misalignment within the employee's state of mind. This affects their PC and their EE within the organization.

Should the employer fail to fulfill their promises, some candidates will feel that their psychological contract has been violated (Aggarwal et al., 2007). Being promised one thing and receiving something else will be difficult for the candidate. This concept is the same for interpersonal influences. This link further demystifies employees' complex behavioral issues, exclusively when interpersonal influences impact the organizational culture.

This research has provided an essential addition to HR practice's accepted features: the Psychological Contract and Employee Engagement paradigm. The interpersonal influences have been added to show that it negatively affects the individual psychological contract, affecting employee engagement. When a person believes that they should receive certain benefits and does not, this creates misalignment and impacts motivation. Moreover, it affects the employee's PC and their EE within the organization.

According to this research, recruitment and psychological contract have a special relationship where, during the recruitment process, the employer makes promises to the new candidate. Should the employer fail to fulfill their promises, some candidates feel that their psychological contract has been violated (Aggarwal et al., 2007). Being promised one thing and having something else given is difficult for the candidate. This concept is applied to interpersonal influences. When a candidate is promised a position *based on* and supported by their so-called 'connections,' their psychological contract is focused on the promise a nd expects a positive outcome.

Each interpersonal influence impacts business. It is also acknowledged that interpersonal influences impact the basic human resource practices through the foundations of human and social capital. These practices steer employees to either develop themselves or not. The model theory is used as the method for this study to enhance how leaders, managers, and HR practitioners view interpersonal influences.

There can be avenues in further research regarding the Individual HR practices concerning $interpersonal influences$; examine the development of HR practices and management; examine the Social Network Analysis HR practices and effects of culture; and HR models that incorporate interpersonal influences.